\documentclass[journal,twoside,web]{ieeecolor}
\usepackage{tmi}
\usepackage{cite}
\usepackage{amsmath,amssymb,amsfonts}
\usepackage{algorithmic}
\usepackage{graphicx}
\usepackage{textcomp}
\usepackage{multirow}
\usepackage{amssymb}
\usepackage{array}
\usepackage{url}
\usepackage{enumerate}
\usepackage{amsmath}
\usepackage{booktabs}
\usepackage{multirow}
\usepackage{amssymb}
\usepackage{amsmath}
\usepackage[linesnumbered, ruled]{algorithm2e}
\usepackage{newfloat}
\usepackage{listings}

\def\etal{\textit{et al.}}

\usepackage{amsmath}
\usepackage{booktabs}
\usepackage{multirow}
\def\BibTeX{{\rm B\kern-.05em{\sc i\kern-.025em b}\kern-.08em
    T\kern-.1667em\lower.7ex\hbox{E}\kern-.125emX}}
\begin{document}
\title{BCNet: Bronchus Classification via Structure Guided Representation Learning}
\author{Wenhao Huang, Haifan Gong, \IEEEmembership{Graduate Student Member, IEEE},  Huan Zhang, Yu Wang, Xiang Wan, Guanbin Li,  \IEEEmembership{Member, IEEE}, Haofeng Li, \IEEEmembership{Member, IEEE}, Hong Shen
\thanks{Haifan Gong and Wenhao Huang contribute equally to this work. Guanbin Li and Haofeng Li are the corresponding authors. 
This work was supported in part by the Shenzhen Science and Technology Program JCYJ20220818103001002), in part by the Guangdong Provincial Key Laboratory of Big Data Computing, The Chinese University of Hong Kong, Shenzhen, in part by the Longgang District Special Funds for Science and Technology Innovation (LGKCSDPT2023002), in part by the National Natural Science Foundation of China (No.62102267), and in part by the Guangdong Basic and Applied Basic Research Foundation (2023A1515011464).}
\thanks{Haifan Gong, Xiang Wan, and Haofeng Li are with Shenzhen Research Institute of Big Data, Shenzhen, 518172, China (e-mail: wanxiang@sribd.cn; lhaof@sribd.cn). Haifan Gong is also with the School of Science and Engineering, The Chinese University of Hong Kong, Shenzhen, 518172, China (e-mail: haifangong@link.cuhk.edu.cn).}
\thanks{Guanbin Li is with the School of Computer Science and Engineering, Sun Yat-sen University, Guangzhou, 510006, China (e-mail: liguanbin@mail.sysu.edu.cn).}
\thanks{Hong Shen, Huan Zhang, Yu Wang, and Wenhao Huang are with InferVision, Beijing, 100000, China (e-mail: shong@infervision.com; zhuan@infervision.com; wyua@infervision.com; hwenhao@infervision.com)}
}

\maketitle

\begin{abstract}
CT-based bronchial tree analysis is a crucial step for the diagnosis of lung and airway diseases. However, the varying topology of bronchial trees across individuals presents a significant challenge to automatic bronchus classification. To address this issue, we propose the Bronchus Classification Network (BCNet), a structure-guided framework that leverages segment-level topological information using point clouds to learn voxel-level features. BCNet consists of two branches: a Point-Voxel Graph Neural Network (PV-GNN) for segment classification and a Convolutional Neural Network (CNN) for voxel labeling. These two branches are trained simultaneously to learn topology-aware features for their shared backbone, while also allowing the CNN branch to be run independently during inference. This ensures that BCNet retains the same inference efficiency as its CNN baseline. Experimental results demonstrate that BCNet significantly surpasses the state-of-the-art methods by over 8.0\% in F1-score for bronchus classification. Additionally, we introduce BronAtlas, an open-access benchmark for bronchus imaging analysis, featuring high-quality voxel-wise annotations of both anatomical and abnormal bronchial segments. The BronAtlas dataset provides a valuable resource for the research community, enabling the development and evaluation of advanced bronchial tree analysis methods for disease diagnosis and surgical planning. The benchmark dataset is publicly available at link\footnote{https://osf.io/pskr9/?viewonly=94fa3d87274b4095ac9a4b88cc9a1341.}. 
\end{abstract}


\begin{figure}[!t]
\begin{center}
\includegraphics[width=1.0\linewidth]{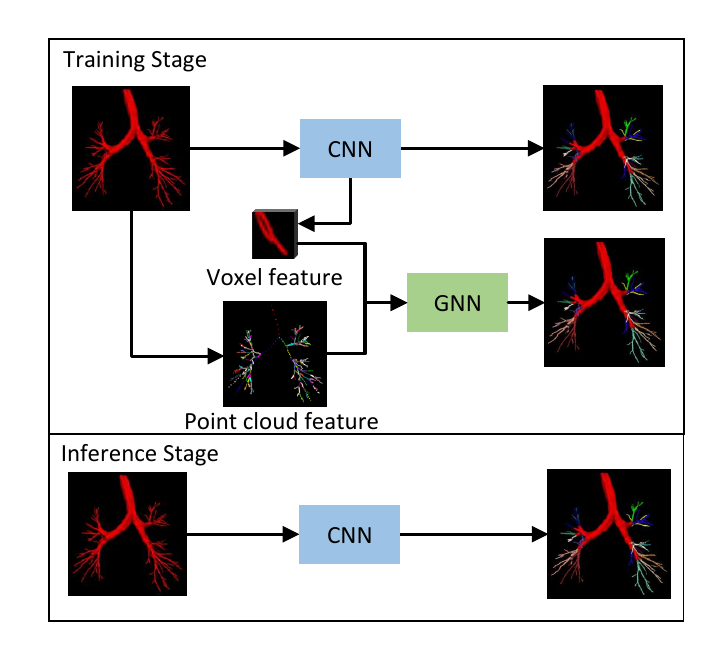}
\end{center}
\caption{The key idea of our BCNet. The upper part shows the training process involves the GNN and CNN. The GNN and CNN parts share the same voxel-wise CNN feature map, which is used to represent the bronchus in the GNN part. By computing loss in the GNN part, we encourage the shared feature map to better describe the bronchus segments and the topological relationships between segments. It can be viewed as additional contextual guidance that is different from and complements the voxel-level supervision in the CNN part. The bottom part shows that our inference phrase only needs the CNN part.} \label{snap}
\end{figure}

\section{Introduction}
Lung airway analysis based on CT imaging is clinically critical as it provides quantitative information that assists in lung disease diagnosis and surgical navigation \cite{ref1,qin2019airwaynet,kawata2023representation}. The reconstruction of the bronchial tree, which serves as the foundation for lung airway analysis, typically involves two key steps \cite{Li2022HumanTT,2p5dnetforairway,meng2017tracking}. The first step involves extracting a binary airway-tree mask from CT images, which delineates the structure of the bronchial tree. The second step is the classification of bronchus segments, where major anatomical branches are labeled accordingly. This automatic reconstruction of the bronchial tree supports various clinical processes, including the matching of individual airway tree phenotypes and the classification of lung lobes or segments \cite{ref6,ref7}. Furthermore, accurate labeling of bronchial segments is crucial for clinicians as it helps identify which anatomical segments are affected by disease, thereby aiding in surgical planning and intervention \cite{mori2000automated,sato2019concepts,keuth2024airway,gong2024intensity}. Despite these advancements, automatic bronchus segment classification still faces a significant challenge: efficiently modeling and exploiting the complex topology of the bronchial tree. Addressing this challenge is essential for improving the accuracy and reliability of bronchial tree reconstruction, ultimately enhancing clinical outcomes.

Motivated by the significance of structural priors in medical image analysis \cite{gong2021multi, gong2023thyroid}, we propose a novel framework, the Bronchus Classification Network (BCNet), designed to enhance the classification of bronchus segmentation from CT images. BCNet capitalizes on the intrinsic topology of the bronchial tree, incorporating a Point-Voxel Graph Neural Network (PV-GNN) to augment the representation learning capabilities of a Convolutional Neural Network (CNN). The core innovation of BCNet lies in its ability to employ segment-wise topological information through the GNN to inform and guide the learning of voxel-wise convolutional features via the CNN. This synergistic approach enables the CNN component to more accurately distinguish bronchus segments at the voxel level, as depicted in Figure~\ref{snap}. Significantly, during inference, BCNet operates solely with the convolutional branch, obviating the need for the graph model. This allows BCNet to achieve the accuracy enhancements afforded by GNN training while preserving the computational efficiency characteristic of a CNN-based approach.

To construct the GNN branch of BCNet, we integrate high-dimensional local features from the CNN with positional and angular information described by point clouds. Recognizing that adjacent bronchus segments often belong to the same category, we introduce a Neighborhood Consistency Regularization as a loss term to enhance model training. For evaluation, we manually annotated airway branches in 100 CT scans, comprising 60 scans from public datasets and 40 from a collaborating hospital. This comprehensive annotation effort ensures robust assessment of both anatomical and abnormal labels.
Overall, our contributions are summarized as follows:
\begin{itemize}
\item BronAtlas, a benchmark containing 100 bronchial cases with accurate voxel-level segmentation masks and anatomical categories including the abnormal bronchus segment;
\item BCNet, a structure-guided representation learning framework to classify the bronchus from lung CT imaging, which significantly outperforms the state-of-the-art methods on BronAtlas; 
\item PV-GNN, a Point-Voxel Graph Neural Network to perform the segment-level bronchus classification, which uses the point cloud and convolutional feature to represent the relative position and semantics of the bronchus segment, respectively. By taking both the convolution feature and point cloud feature as the node feature to construct the graph, the PV-GNN has shown its superiority through ablation studies.
\end{itemize}



\section{Related Work}
\subsection{Bronchus Analysis Based on Deep Learning}
In recent years, deep learning (DL) has significantly advanced bronchus segmentation \cite{Li2022HumanTT,zhang2023multi}. Various single-stage networks, such as U-Net \cite{ronneberger2015u}, 2.5D Net \cite{2p5dnetforairway,heitz2021lubrav}, and 3D U-Net \cite{meng2017tracking,charbonnier2017improving,isensee2021nnunet,qin2019airwaynet,qin2020learning,qin2021tscnn,garcia2021automatic,zheng2021alleviating,gong2024intensity}, have been employed for this task. Meng \etal \cite{meng2017tracking} proposed an image tracking method to enhance segmentation accuracy. Qin \etal \cite{qin2020learning, qin2021tscnn} developed CNN-based methods that demonstrate superior sensitivity to tenuous peripheral bronchioles. Zheng \etal \cite{zheng2021alleviating} applied gradient erosion and dilation operators to mitigate inter-class imbalance issues in bronchus segmentation, while their subsequent work \cite{zheng2021refined} introduced gradient ratio adjustment and weight enhancement strategies to handle hard-to-segment regions. Tang \etal \cite{tang2023adversarial} introduced an adversarial transformer network for improved airway segmentation, and Nan \etal \cite{nan2023fuzzy} developed a fuzzy attention neural network aimed at addressing segmentation discontinuities. Additionally, Gong \etal \cite{gong2024intensity} proposed an intensity-clarified loss function to better segment the bronchus from CT scans.

Bronchus classification \cite{tschirren2005matching,feragen2012hierarchical,Li2022HumanTT} remains a challenging task due to the substantial variability in bronchial tree topology across individuals. This classification typically builds on the results of bronchus segmentation methods \cite{heitz2021lubrav,charbonnier2017improving,isensee2021nnunet,qin2019airwaynet,qin2020learning,qin2021tscnn,garcia2021automatic,zheng2021alleviating,zheng2021refined, tang2023adversarial}. Traditional methods for bronchus classification include geodesic matching \cite{feragen2012hierarchical} and probabilistic hypergraph matching \cite{liu2022automated}. Wang \etal \cite{ref7} achieved lobar-level bronchus classification using a key-point detection approach, while Lo \etal \cite{lo2011bottom} proposed a bottom-up rule-based method for bronchial labeling. In the realm of deep learning-based methods, Zhao \etal \cite{bronchuslp-zhao-2019} conducted an analysis of the average inclination angle of bronchial segments in the training set and utilized linear programming to post-process the airway structures predicted by neural models. Nadeem \etal \cite{ref6} introduced a two-stage neural network that labels the lobar-level and segment-level bronchus respectively, ensuring detailed and hierarchical classification. Further, Yu \etal \cite{yu2022tnn} proposed the tree neural network for comprehensive airway labeling, demonstrating the efficacy of structured neural networks in capturing complex bronchial topologies.

\subsection{Graph Neural Networks for Segmentation}
Recently, Graph Neural Networks (GNNs)~\cite{gcn,2018gat,gin} have become increasingly important in the biomedical domain~\cite{song2022diagnosis,li2023hierarchical,liang2022risk,gong2023unbiased,corso2024graph}, particularly for addressing segmentation tasks \cite{garcia2019joint,selvan2020graph,zhao2021airway,tan2021sgnet,lou2024structure,lou2024cell}. For instance, Garcia-Uceda \etal \cite{garcia2019joint} enhanced airway binary segmentation by replacing the deepest convolutional layer in a U-Net \cite{ronneberger2015u} with graph convolutions \cite{gcn}, demonstrating the potential of GNNs in improving segmentation results. Similarly, Selvan \etal \cite{selvan2020graph} introduced a GCN-based \cite{gcn} mean-field network to refine segmentation outputs produced by a 3D-UNet \cite{ronneberger2015u}, further highlighting the role of GNNs in segmentation refinement. Tan \etal \cite{tan2021sgnet} approached bronchial segmentation and classification as a semantic segmentation problem, proposing a two-stage framework specifically designed for bronchus classification. Zhao \etal \cite{zhao2021airway} developed a prototype-based graph neural network aimed at detecting abnormal bronchus structures, showcasing the application of GNNs in identifying pathological changes.

However, these prior methods often rely heavily on hand-crafted features \cite{ref7,bronchuslp-zhao-2019,ref6} or require additional labels \cite{tan2021sgnet,garcia2019joint}, which can limit their adaptability and efficiency. Moreover, they frequently struggle to manage the individual variability in bronchial structures effectively, which is crucial for accurate classification and segmentation. To address these limitations, there is a pressing need for an efficient framework that leverages both voxel-wise features and point-cloud topology for bronchus classification. Our proposed Bronchus Classification Network (BCNet) integrates the structural topology of the bronchial tree using a Point-Voxel Graph Neural Network (PV-GNN) to enhance the representation learning capabilities of a Convolutional Neural Network (CNN), thereby improving voxel-level feature discrimination and bronchus classification performance while maintaining computational efficiency.

\section{BronAtlas: A Benchmark for Bronchus Analysis}
\begin{figure}[!t]
\begin{center}
\includegraphics[width=1.0\linewidth]{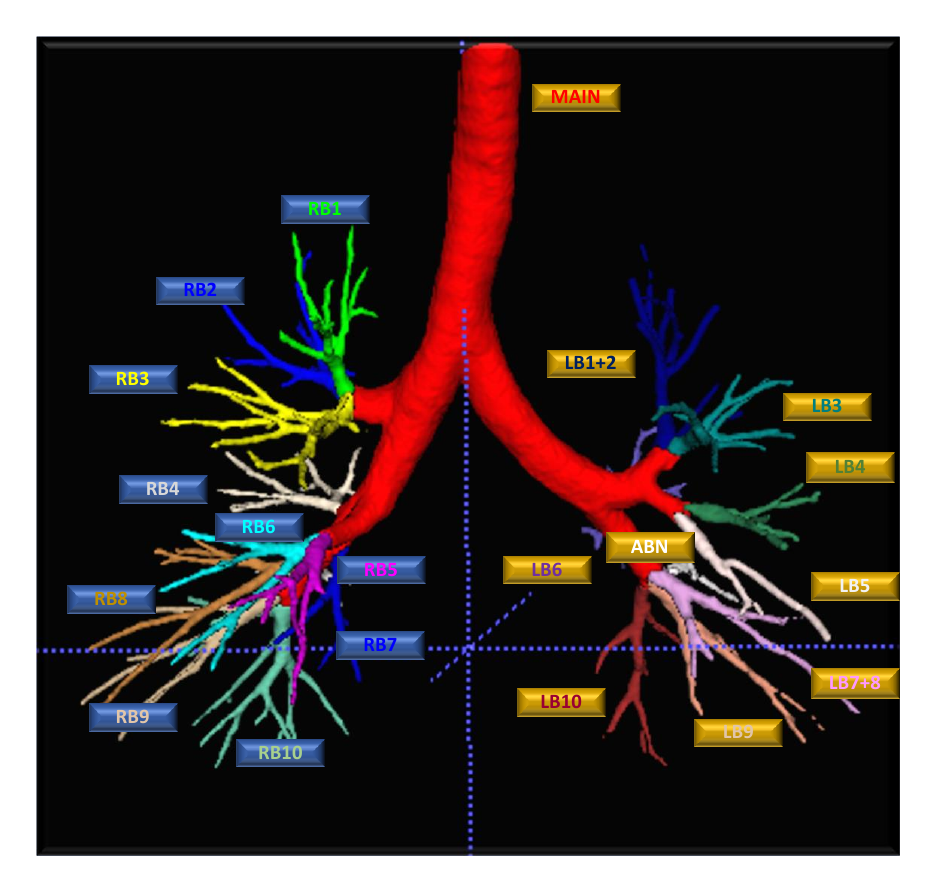}
\end{center}
\caption{The visualization results of the bronchus segment labels. The segmental bronchi from the right lung (i.e., RB1 to RB10) and segmental bronchi from the left lung (i.e., LB1+2, LB3, ..., LB7+8, LB9, LB10) are defined according to \cite{tschirren2005matching}. ``MAIN'' denotes the main trachea while ``ABN'' denotes the abnormal bronchus segment. The radiologists use the binary segmentation of the bronchus and make sure the bronchial segments are connected to the main bronchial trunk except for the abnormal branches. For the bronchus segments that do not belong to the above classes, the radiologists regard them as abnormal bronchus.} \label{label}
\end{figure}

\begin{table}[h]
\centering
\caption{Details of the BronAtlas: a benchmark for bronchus classification with a voxel-wise mask. We collect 60 cases from the currently available databases EXACT'09~\cite{EXACT} and LIDC~\cite{lidc}. The remaining 40 cases are collected from our cooperative hospital. For the currently public samples and the newly collected ones, each CT scan is annotated by two experts in a two-step annotation process. The experts first annotate binary masks for the airway segmentation, and then label 20 segmental bronchi at the voxel level according to the anatomical structure in Figure~\ref{label}.}
\begin{tabular}{cc}
\hline
Attributes of BronAtlas             & Values          \\ \hline
Resolution of axial slices          & 512 $\times$ 512 \\
Thickness of axial slices & 0.55 $\sim$ 1.00 mm \\
\#Samples in BronAtlas & 100 \\
\#Samples from LIDC & 40 \\
\#Samples from EXACT'09 & 20 \\ 
\hline
\end{tabular}
\label{dataset_summary}
\end{table}

\begin{figure*}[!t]
\begin{center}
\includegraphics[width=1.0\textwidth]{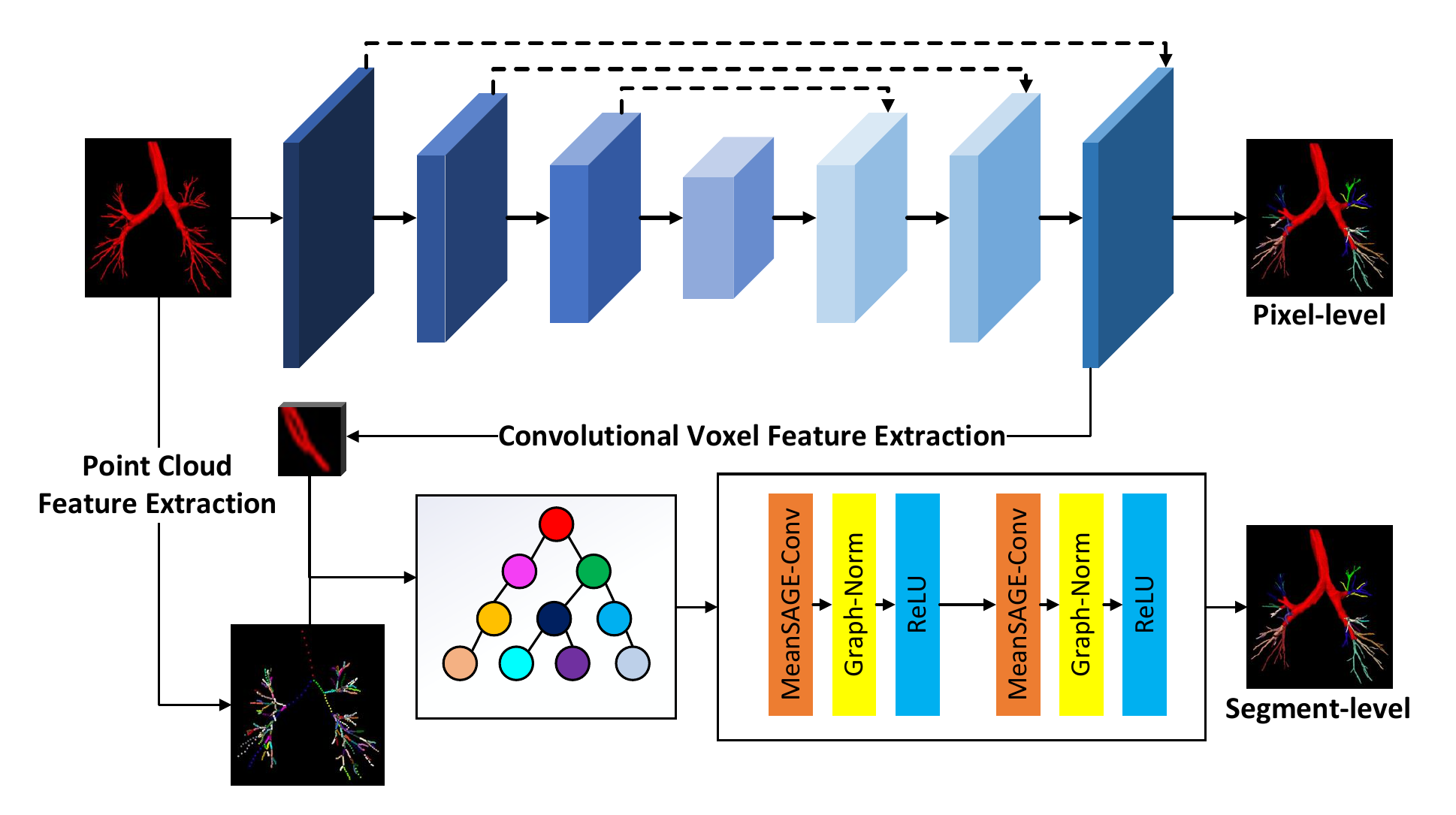}
\end{center}
\caption{Overview of the Bronchus Classification Network (BCNet). The upper part depicts a UNet that predicts voxel-level categorical labels of the bronchus based on a binary bronchus mask, refining segmentation through multiple convolutional layers. The lower part illustrates a Graph Neural Network (GNN) that operates at the segment level, predicting bronchus categories by leveraging structural and topological information. During training, gradients from the GNN are propagated backward through the UNet, guiding its representation learning and enhancing overall classification accuracy.} \label{network:bcnet}
\end{figure*}

\subsection{Motivation}
The category and the abnormality of bronchus segments play an important role in computer-aided bronchial disease diagnosis. However, the existing bronchial classification dataset does not contain the categorical label of abnormal bronchial segments. 
Moreover, we observe that the existing bronchial segmentation datasets \cite{EXACT,lidc,qin2019airwaynet} do not provide the category of bronchial segments. To facilitate the development of automatic bronchus diagnosis, we contribute a new benchmark, BronAtlas, for bronchus segmentation and classification. BronAtlas contains 100 cases of lung CT imaging with voxel-level annotations of 20 segmental bronchi categories (i.e., 10 from the right lung, 8 from the left lung, 1 for the main trachea, and 1 for abnormal bronchus). We will make this dataset available after acceptance.

\subsection{Sample Collection and Annotation}
We collected 60 cases from the publicly available databases EXACT'09 \cite{EXACT} and LIDC \cite{lidc}, and an additional 40 cases from our cooperative hospital. For both the public samples and the newly collected ones, each CT scan was annotated by two experts through a two-step annotation process. Initially, the experts created binary masks for airway segmentation. Subsequently, they labeled 20 segmental bronchi at the voxel level based on the anatomical structure, as illustrated in Fig.~\ref{label}. 

Following the binary segmentation of the bronchus, the radiologists labeled 18 bronchial segments and the main trachea according to \cite{tschirren2005matching}. Specifically, 10 segmental bronchi were labeled in the right lung (RB1 to RB10), and 8 in the left lung (LB1+2, LB3, ..., LB7+8, LB9, LB10). During the annotation process, the radiologists ensured that, except for the abnormal branches, all bronchial segments were connected to the main bronchial trunk and that there were no issues of misclassification, under-classification, non-abnormal co-drying, or cross-color. Bronchial segments not fitting these categories were classified as abnormal bronchi. The process of data acquisition and annotation adhered to the principles outlined in the Declaration of Helsinki \cite{williams2008declaration} and received approval from the institutional ethical committee.

\subsection{Dataset Summary}
For the development of automatic bronchus diagnosis, we introduce a new benchmark, BronAtlas, specifically designed for bronchus analysis. BronAtlas comprises 100 cases of lung CT imaging, each with voxel-level annotations for 20 segmental bronchi categories (10 from the right lung, 8 from the left lung, 1 for the main trachea, and 1 for abnormal bronchus). The ``abnormal bronchus segment'' refers to congenital abnormal bronchus. The dataset includes 33 abnormal cases in the training and validation set, and 12 abnormal cases in the test set. A visualization of the 20 bronchial categories is provided in Figure~\ref{label}, and a detailed summary of the samples in the benchmark is presented in Table~\ref{dataset_summary}. The data from different sources were combined and split into training, validation, and test sets with 63, 7, and 30 cases, respectively.

The patient data used in this study was anonymized, with all personally identifiable information removed prior to model training and evaluation to ensure privacy. To minimize bias, we utilized a diverse dataset that includes CT scans from individuals of various ages, genders, and health conditions. Despite these efforts, we acknowledge the need for a more systematic approach to addressing bias, such as stratified sampling or bias audits, which we aim to implement in future work. This comprehensive and meticulously curated dataset aims to facilitate advancements in the automatic diagnosis of bronchial conditions.

\begin{figure*}[!t]
\begin{center}
\includegraphics[width=1\linewidth]{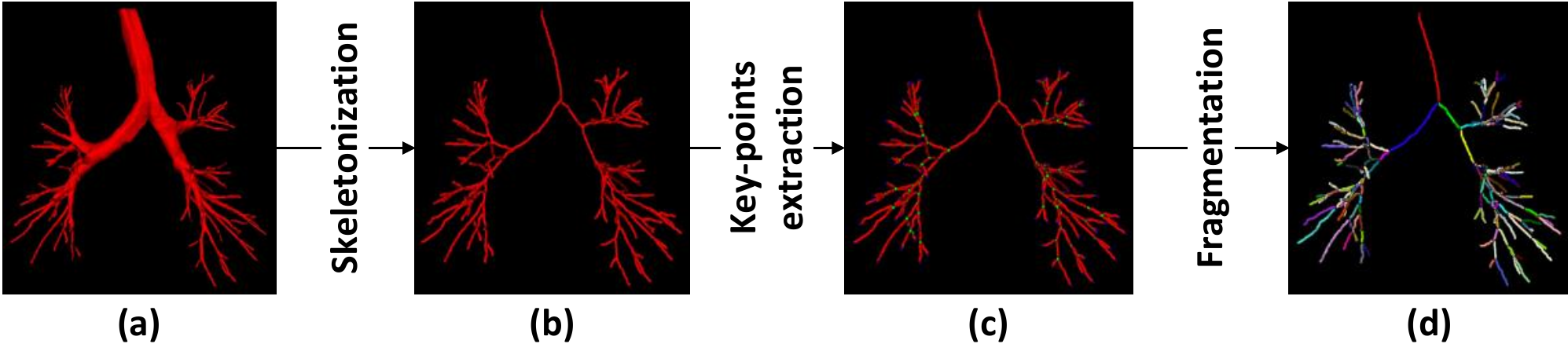}
\end{center}
\caption{Construction of bronchial tree: (1) skeletonize the bronchus; (2) extract the key points separating the bronchial fragments; (3) separate the bronchus into segments.} \label{btc}
\end{figure*}

\section{Method}
To use the inherent topology of the bronchus to guide the representation learning for bronchus classification, we proposed the Bronchus Classification Network (BCNet), which is shown in Fig.~\ref{network:bcnet}. The proposed BCNet includes two components: a UNet for voxel-level bronchus labeling, and a Point-Voxel Graph Neural Network (PV-GNN) for segment-level bronchus labeling. The structure-guided representation learning in BCNet means that the UNet can benefit from the gradients propagated from the PV-GNN branch since the PV-GNN adopts the convolutional feature from the UNet branch to construct graph nodes.

\subsection{Motivation}
We employ the UNet to obtain the voxel-level bronchus segmentation result since the UNet can compute the high-dimension voxel feature for the following segment-level bronchus labeling. The motivation behind our proposed GNN is that the relative positional information represented by the point cloud can help model the intrinsic bronchus topology during the classification process. In the meanwhile, the high-dimension voxel feature could be a strong supplement for bronchus classification since it can implicitly capture other kinds of bronchus attributes such as the diameter of bronchi. Finally, we utilize the GNN to guide the learning process of the UNet through the convolution feature and improve the classification performance of the UNet. The joint learning of the GNN and CNN branches can be seen as a form of regularization. By encouraging the model to learn representations on two tasks, the model can achieve better generalization performance. Note that the gradients of GNN loss are passed through the CNN, while those of CNN loss are not sent into the GNN. Thus, the CNN is trained with the voxel-level loss and an additional segment-level loss via the feature construction procedure of GNN, while the GNN part is only trained with the segment-level loss. The CNN can learn from both tasks at the voxel level and segment level and exploit the shared information to improve performance without introducing additional parameters.

\subsection{UNet for Voxel-level Bronchus Labeling}
We classify the bronchus at voxel level with a UNet~\cite{isensee2021nnunet} that is shown in the upper part of Fig.~\ref{network:bcnet}. The UNet can produce both the coarse bronchial prediction and the voxel-level feature for further segment-level classification. The UNet is trained with the cross-entropy loss, which is formulated as: 
\begin{equation}\label{eq:unetloss}
    L_{unet} = L_{CE}(y_{pred}, y_{gt}),
\end{equation} 
where the ground truth and the prediction are denoted as $y_{gt}$ and $y_{pred}$, respectively. It is worth noting that the background class is not considered.

\subsection{Point-Voxel GNN for Segment-level Bronchus Labeling}
\paragraph{Definition of Point-Voxel Graph}
Given a binary segmentation mask of the bronchus, we update the mask as its maximum connected region and define a point-voxel graph with the region, 
using mixed voxel and point-level features as node embeddings. The point-voxel graph was built, as shown in Fig.~\ref{btc}. 
First, we skeletonize the binary mask by extracting the centerline (see Fig. \ref{btc}(b)) via an existing algorithm \cite{lee1994building}. Second, according to the number $N$ of foreground voxels in the 26-connected neighborhood of each voxel on the centerline, we define the endpoints ($N=1$), edge points ($N=2$), and division points ($N \geq 3$), as shown in Fig.~\ref{btc}(c). Finally, as Fig.~\ref{btc}(d) displays, we divide the branches into segments with these points. Each bronchial segment corresponds to a node in the proposed point-voxel graph. An edge of the graph is defined by the connectivity between two line segments that are divided by a division point. The details of our network structure are shown in Table~\ref{tab:net-structure}.

\begin{table*}[!ht]
\centering
\caption{Details of the UNet and Point-Voxel Graph Neural Network (PV-GNN) structures and parameters. ``BN'' denotes batch normalization, ``SK'' denotes skip connection, ``FC'' denotes fully connected layer, ``Conv'' denotes convolution, ``MSC'' denotes Mean Sage-Convolution, ``GN'' denotes Graph Normalization, and ``Conv-Norm Block'' combines MSC, GN, and ReLU layers. ``K'' denotes the number of points per node in the graph.}
\label{tab:net-structure}
\begin{tabular}{@{}lllll@{}}
\toprule
\textbf{Module} & \textbf{Input Size/Channels} & \textbf{Network Layer/Operation} & \textbf{In/Out Channels} & \textbf{Description} \\ \midrule
\multicolumn{5}{c}{\textbf{UNet Architecture}} \\ \midrule
Encoder & $128 \times 128 \times 128$ & Stem Conv & 1 / 24 & Conv3d+BN+ReLU \\
        & $64 \times 64 \times 64$ & Residual block 1 & 24 / 48 & Multiple layers with SK \\
        & $32 \times 32 \times 32$ & Residual block 2 & 48 / 96 & Multiple layers with SK \\
        & $16 \times 16 \times 16$ & Residual block 3 & 96 / 192 & Multiple layers with SK \\
Decoder & $32 \times 32 \times 32$ & Upsample layer & 192 / 192 & Trilinear upscaling \\
        & $32 \times 32 \times 32$ & Double Conv & 192+96 / 96 & Conv3d+BN+ReLU \\
        & $64 \times 64 \times 64$ & Upsample layer & 96 / 96 & Trilinear upscaling \\
        & $64 \times 64 \times 64$ & Double Conv & 96+48 / 48 & Conv3d+BN+ReLU \\
        & $128 \times 128 \times 128$ & Upsample layer & 48 / 48 & Trilinear upscaling \\
        & $128 \times 128 \times 128$ & Double Conv & 48+24 / 24 & Conv3d+BN+ReLU \\
        & $128 \times 128 \times 128$ & FC layer & 24 / 20 & Reduce dimensions \\
\midrule
\multicolumn{5}{c}{\textbf{PV-GNN Architecture}} \\ \midrule
PV-GNN & $(24 + 3) \times K$ & GNN Block 1 & $(24 + 3) \times K$ / $(24 + 3) \times K$ & Conv-Norm Block \\
       & $(24 + 3) \times K$ & GNN Block 2 & $(24 + 3) \times K$ / $(24 + 3) \times K$ & Conv-Norm Block \\
       & $(24 + 3) \times K$ & GNN Block 3 & $(24 + 3) \times K$ / $(24 + 3) \times K$ & Conv-Norm Block \\
       & $(24 + 3) \times K$ & Fully-connected layer & $(24 + 3) \times K$ / 20 & Reduce dimensions \\
\bottomrule
\end{tabular}
\end{table*}

\paragraph{Point-wise Coordinate Feature}
We generate the point cloud from the bronchus mask to obtain the point-wise coordinate feature.
We crop out the bounding box of each bronchial segment (i.e., a node in the graph) from the bronchus tree. Each segment is composed of the centerline voxels that are within the bounding box. Then the coordinate ($X$, $Y$, and $Z$) of each voxel is normalized to [0, 1] concerning the shape of the bounding box. Such normalization is expected to help describe the pose and angle of a segment. Let $L$ be the length of the center line, We divide the centerline into $K-1$ parts with the same interval $\frac{L}{K-1}$, and collect $K$ points from the start and the end points of these $K-1$ parts. 
For each bronchial segment, we define its \textit{point-wise coordinate feature} as the list of three-dimensional coordinates of these $K$ points. In this way, for each bronchial segment, we obtain its point cloud feature that contains three-dimensional coordinates of $K$ voxels and has a length of $3 \times K$. The detailed analysis of the hyper-parameter $K$ is shown in Table~\ref{tab:ablacls}. 

\paragraph{Voxel-wise Convolution Feature}
To obtain voxel features, we train a UNet (the upper part in Fig.~\ref{network:bcnet}) to predict the category of each bronchus branch. A 3D feature map is produced from the penultimate layer of the UNet. For each bronchial fragment, we locate the $K$ voxels that are nearest to the above-mentioned $K$ points and extract the convolution features from the feature map according to each voxel's indices. The \textit{voxel-wise convolution feature} of the bronchial fragment is defined as the combination of the $K$ features and has a length of $C \cdot K$, 
where $C$ is the channel number of the last but one feature map ($C=24$ by default) of the UNet. Thus, the size of a voxel-wise convolution feature is $24 \times K$. 


\begin{table*}[!t]
\centering
\caption{Comparison of bronchus classification methods. LP \cite{bronchuslp-zhao-2019}, TS-NN \cite{ref6},  SGNet \cite{tan2021sgnet} are the previous state-of-the-arts methods. The GNN-based methods (i.e., GCN \cite{gcn}, GAT \cite{2018gat}, GIN \cite{gin}) utilize \textbf{our proposed point-voxel feature} for graph construction. \textit{p-value} is calculated between our model and other models under three different seeds.}
\begin{tabular}{ccccccccc}
\hline
Methods & Accuracy  & Precision & Recall & F1-score       &\textit{p-value} &Flops (G)& Running Time (s)&Throughput (samples/s)\\ \hline
LP \cite{bronchuslp-zhao-2019}& $81.8_{\pm4.3}$               & $74.7_{\pm3.4}$               & $79.2_{\pm4.8}$               & $77.0_{\pm4.2}$                & \textless  0.001 & -& 0.75&1.33\\
TS-NN\cite{nadeem2020ct}              & $76.8_{\pm5.3}$               & $77.8_{\pm4.4}$               & $74.9_{\pm6.1}$               & $76.2_{\pm5.2}$                & \textless  0.001 & 0.38& 0.04&24.99\\
SGNet\cite{tan2021sgnet}& $85.6_{\pm5.3}$ & $85.0_{\pm5.2}$ & $84.4_{\pm6.2}$  & $84.7_{\pm5.6}$   & \textless  0.001 & 477.47& 4.27&0.28\\ \hline
GCN\cite{gcn} & $90.7_{\pm 5.7}$ & $90.2_{\pm 4.6}$ & $90.4_{\pm 5.6}$ & $90.4_{\pm 5.2}$  & 0.006 & 13.21 & 2.52 & 0.40 \\
GIN\cite{gin} & $90.0_{\pm 4.2}$ & $89.2_{\pm 5.3}$ & $90.1_{\pm 5.8}$ & $89.4_{\pm 5.5}$  & \textless  0.001 & 13.12 & 2.51 & 0.40 \\
GAT\cite{2018gat} & $91.2_{\pm 5.6}$ & $91.6_{\pm 4.9}$ & $90.7_{\pm 5.4}$ & $91.1_{\pm 5.2}$  & 0.020 & 14.21 & 2.60 & 0.39 \\ \hline
BCNet                    & \textbf{${94.2_{\pm3.8}}$} & \textbf{${93.4_{\pm6.3}}$} & \textbf{${94.0_{\pm4.9}}$} & \textbf{${93.7_{\pm5.5}}$}  & - & 13.07 & 1.00 & 1.00\\ \hline
\end{tabular}
\label{sota-cls}
\end{table*}

\paragraph{Point-Voxel  Graph  Neural  Network}
Given a point-voxel graph that takes both point-cloud features and high-dimension voxel features into account, we design a Point-Voxel Graph Neural Network (PV-GNN) to predict the category of each bronchial segment. The PV-GNN consists of Conv-Norm Blocks and a fully connected layer. A Conv-Norm Block is composed of a Mean Sage-Convolution (MSC) \cite{msc} layer, a Graph Normalization (GN) \cite{graphnorm} layer, and a ReLU activation function. The MSC layer uses the mean aggregated function to merge information from node neighbors to overcome the inductive bias. Considering that adjacent nodes in the topology of the bronchial tree are relatively sparse, we build up a deep GNN for better information integrating point clouds. Since the GNN has a risk of suffering from gradient vanishing as it goes deeper, we introduce GN to shift and scale feature values, which makes graph neural networks converge much faster. Besides the first block, each block is followed by an element-wise addition that acts as a residual connection. Let $H^k$ denote the output of the $k$-th block and $\sigma$ denote the ReLU operation. The Conv-Norm block is formally defined as Eq.~(\ref{eq:convnorm}):
\begin{equation}\label{eq:convnorm}
H^k = \sigma({\rm{GN}}({\rm{MSC}}(H^{k-1}))+ H^{k-1}.
\end{equation}

\paragraph{CE Loss with Neighborhood Consistency Regularization}
Considering the topology prior that adjacent branches in a bronchial tree more likely belong to the same category, we design a novel Neighborhood Consistency Regularization (NCR) to penalize local spatial variations and force the nearby nodes of the same category to be closer in a latent space. Let $Y = \{y_1, y_2, ..., y_N\}$ be the set that contains the one-hot vector of the ground truth of each branch and $Z = \{z_1, z_2, ..., z_N\}$ be the set that contains the logit vector of the model prediction of each branch. The NCR loss is formulated as Eq.~(\ref{eq:NCR}):
\begin{equation}\label{eq:NCR}
L_{NCR} = \frac{\sum_{i=1}^N \sum_{j}^{V_i} ||z_i - z_j|| \mathbb I (y_i = y_j)}{M},
\end{equation}
where $V_i$ is the set of the $i$-th node's neighbors, $j$ denotes the $j$-th node in this set, $z_i$ denotes the logit vector of the $i$-th node from the output of the last fully connected layer, $\mathbb I(\cdot)$ is an indicator function that returns 1 when the condition is met and returns 0 otherwise, and $M, N$ are the numbers of edges and nodes in the graph, respectively. Let $\alpha$ be a scalar to balance the weight of the regularization and CE loss, the overall loss function contains the above NCR and a vanilla Cross-Entropy loss and is formulated as Eq.~(\ref{eq:cls-NCR}):
\begin{equation}
L_{pvgnn} = L_{CE}(y_{pred}, y_{gt}) + \alpha L_{NCR}(y_{pred}),
\label{eq:cls-NCR}
\end{equation}
where $L_{pvgnn}$ has the same number of classes with $L_{unet}$.

\subsection{Overall Loss for Bronchus Classification}
We train our proposed model with the UNet and PV-GNN components in an end-to-end manner, since the gradients of the GNN branch can sent backward to the UNet branch through the operation of convolutional voxel feature extraction. Let $\beta$ be a scalar to balance the weight of two tasks, the final loss $L_{cls}$ is composed of two parts, which could be formulated as Eq.~(\ref{eq:clsloss}):
\begin{equation}\label{eq:clsloss}
L_{cls} = L_{unet} + \beta L_{pvgnn}.
\end{equation}

\begin{table}[!h]
\centering
\caption{Five-fold cross-validation result of three representative methods. In this setting, we use 56 samples for training, 14 samples for validation, and 30 samples for testing.}
\begin{tabular}{cccccc}
\toprule
Method &Accuracy  & Precision  & Recall  & F1-score  & p-value\\ \midrule
CNN & 81.7$_{\pm0.3}$ & 82.3$_{\pm0.2}$ & 81.1$_{\pm0.4}$ & 81.7$_{\pm0.4}$ & \textless  0.001 \\
GIN & 87.6$_{\pm0.1}$ & 87.1$_{\pm0.2}$ & 86.3$_{\pm0.1}$ & 86.7$_{\pm0.2}$ & \textless  0.001 \\
BCNet & 93.2$_{\pm0.1}$ & 93.0$_{\pm0.1}$ & 92.6$_{\pm0.1}$ & 92.8$_{\pm0.1}$ & - \\
\bottomrule
\end{tabular}
\label{tab:my_label}
\end{table}

\begin{table}[!h]
\centering
\caption{Detailed performance analysis on the out-of-domain data from another data center which contains 30 cases. \textit{p-value} is calculated between our model and other models under three different seeds.}
\begin{tabular}{cccccc}
\toprule
Method &Accuracy  & Precision  & Recall  & F1-score  & p-value\\ \midrule
CNN & 84.5$_{\pm7.0}$ & 83.1$_{\pm7.1}$ & 86.8$_{\pm6.3}$ & 84.9$_{\pm6.7}$ & \textless  0.001 \\
GIN & 88.7$_{\pm0.6}$ & 88.0$_{\pm5.7}$ & 91.4$_{\pm5.4}$ & 89.6$_{\pm5.4}$ & \textless  0.01 \\
BCNet & 91.7$_{\pm4.6}$& 90.3$_{\pm4.9}$& 94.3$_{\pm3.5}$& 92.3$_{\pm4.1}$& - \\
\bottomrule
\end{tabular}
\label{tab:ood}
\end{table}

\section{Experiments and Results}
\subsection{Implementation Details}
PyTorch 1.10 is used to build the model. All models are trained with an NVIDIA V100 GPU of 32GB. BronAtlas is split into the training/validation/testing set with 63/7/30 samples. We use the validation set to tune our parameters select the best-performed model and further test it on the test set. We augment the training data by applying random affine transformation and elastic deformation 99 times. Following the aforementioned data augmentation process, we transformed the samples into a cube with a shape of $128\times 128\times 128$. This standardized cube was used to train all the algorithms in this work. For the testing phase, we reverted the cube to its original shape for evaluation and visualization. By doing this, we were able to ensure a fair comparison across all the methods. We use DropEdge \cite{dropedge} for the classification model training to avoid over-fitting. The model is trained with the Adam optimizer, a batch size of 128, and a learning rate of 0.001 for 50 epochs. The number of layers and hidden dimensions in PV-GNN is set to 5 and 256, respectively. 

\subsection{Evaluation Metrics}
We follow ~\cite{bronchuslp-zhao-2019} to adopt four metrics: accuracy, precision, recall, and F1-score. The overall F1-score is computed by taking the average of the F1-scores of all classes, while the F1-score of each class is based on the harmonic mean of the precision and recall. All performance metrics reported herein are computed using the One-versus-All (OvA) strategy. OvA treats one class as the positive class and combines all other classes into a single negative class for each classification task, thereby enabling the evaluation of classifier performance on a per-class basis.

\subsection{Comparison with the State-of-the-art Methods}
The quantitative comparison for bronchial classification is shown in Table~\ref{sota-cls}. The previous state-of-the-art methods, we implement them by following the description in their paper and using the same training setting as our work. For the GNN-based methods, we first pre-train a UNet to segment the bronchus and extract the dimension voxel feature, then use the GNN that takes the point cloud feature and high-dimension voxel feature as the node feature for bronchus classification. Our final BCNet is trained using a multi-task learning pipeline with the PV-GNN and CNN. For inference, BCNet solely relies on the CNN part.

\begin{table}[!h]
\centering
\caption{Performance on the abnormal category in bronchus atlas benchmark. Results are tested using the best-performing model in the validation set.}
\begin{tabular}{cccccc}
\toprule
Method &Accuracy  & Precision  & Recall  & F1-score &  p-value  \\ \midrule
CNN       & 24.5 & 26.7 & 15.4 & 19.5 & \textless  0.001  \\
GIN       & 39.3 & 38.2 & 22.0 & 27.9 & \textless  0.001 \\
BCNet & 61.2 & 60.7 & 50.0 & 54.8 & - \\ \bottomrule
\end{tabular}
\label{tab:abnormal}
\end{table}

Observations from Table~\ref{sota-cls} indicate that our uniquely designed point-voxel feature allows the Graph Neural Networks (GNNs) to deliver superior results compared to previously established state-of-the-art methods. This shows that compared with the previous handcraft feature \cite{bronchuslp-zhao-2019} or other graph construction feature \cite{tan2021sgnet}, the point cloud feature and convolutional feature are quite important for the distinguishable graph representation learning.
Our Bronchus Classification Network (BCNet) significantly outperforms the previously established state-of-the-art SGNet, achieving an improvement of 8.8\% in accuracy and 9.0\% in the F1-score. This underscores the substantial effectiveness of our proposed structure-guided learning framework.
Notably, BCNet exhibits an advantage in efficiency over Graph Neural Network (GNN)-based methods. While the latter requires both Convolutional Neural Network (CNN) and GNN components during the inference stage, BCNet only needs the CNN part for inference, leading to greater efficiency and compromising performance. We present the outcomes of representative methods in Table~\ref{tab:my_label}, conducted under a five-fold cross-validation setting on our dataset. To achieve five-fold cross-validation, we adopt the data partition of ``56/14/30''. Note that, for the data partition of ``63/7/30'' in Section III.C, we do not perform cross-validation. Additionally, we further evaluate our model's performance on out-of-domain data, as shown in Table~\ref{tab:ood}. All these results support that our methods are effective.

\begin{figure*}[!t]
\begin{center}
\includegraphics[width=\textwidth]{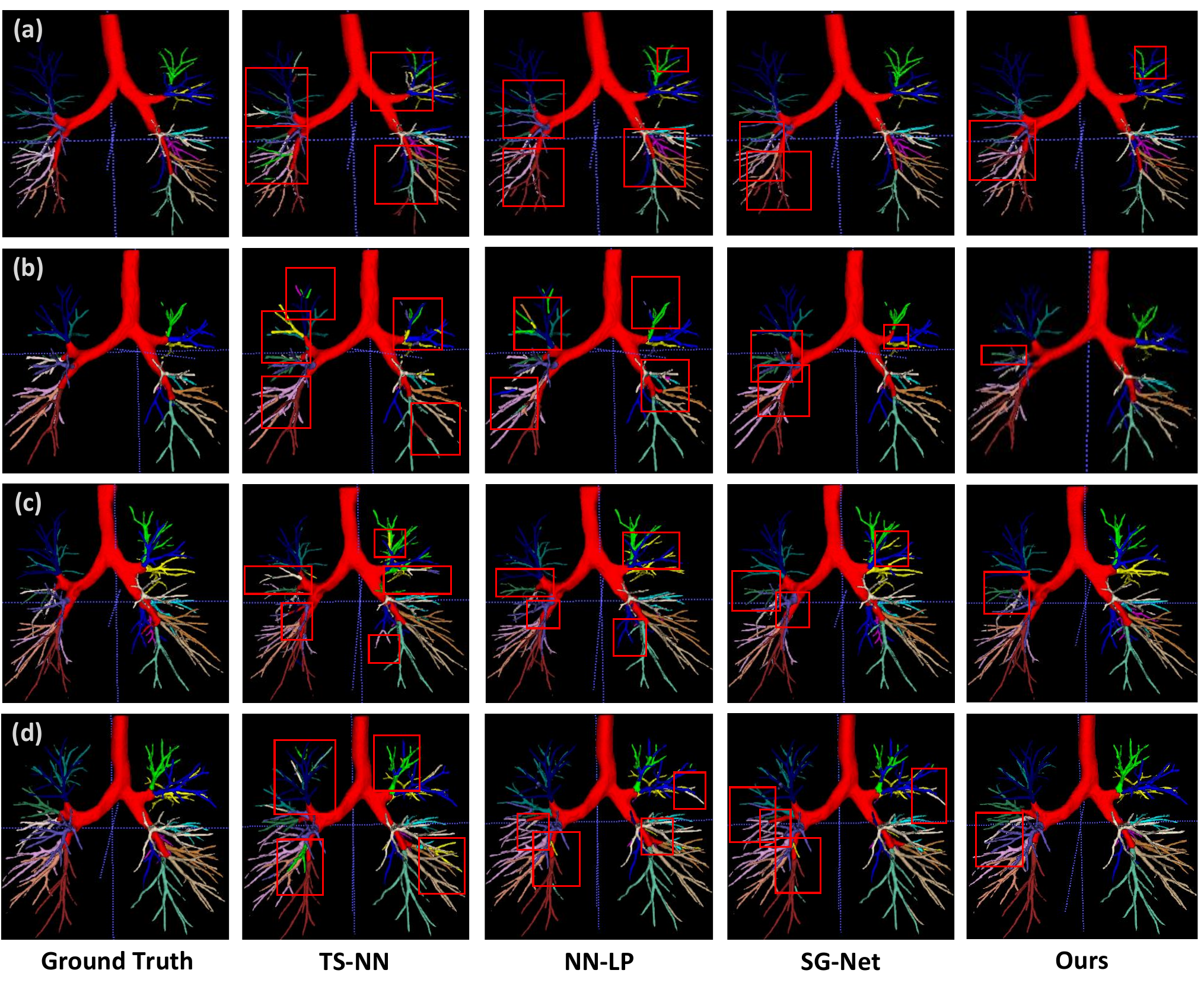}
\end{center}
\caption{Qualitative analysis on the BronAtlas benchmark. The misclassified bronchus is bounded by red boxes. TS-NN \cite{ref6} suffers from errors related to branching variability. LP \cite{bronchuslp-zhao-2019} fails to distinguish the segments with similar angles. SGNet \cite{tan2021sgnet} misclassifies the thin bronchus while the proposed BCNet shows robust predictions.}
\label{visualization}
\end{figure*}

We also provide the averaging result of the abnormal cases, and it is shown in Table~\ref{tab:abnormal}. We can see that our method achieves the highest performance and significantly exceeds the previous methods on the abnormal samples by over 25\% on the F1 
scores. The abnormal cases are clinically important. However, due to the limited abnormal cases in the training set, there is still a gap to achieve better performance, which will be our future work.

\subsection{Ablation Study}
The ablation study for the classification task is in Table~\ref{tab:ablacls}. `CNN' uses only a UNet to label the input binary mask, which shows inferior results due to the lack of structure modeling. `*-GNN' denotes the method that adopts a graph neural network with different node representations. 
`PF' denotes the point-wise coordinate feature based on point clouds, while `VF' means the voxel-wise convolution feature. `NCR' denotes Neighborhood Consistency Regularization. 
Using the graph representation of point clouds, `P-GNN' outperforms the `CNN' by 6.3\% on the F1-score. 
`PV-GNN' means the proposed point-voxel graph neural network that adopts both the voxel-wise convolutional features and the point cloud features as the node representations, and it can capture the texture and diameter information, thus significantly exceeding the `P-GNN' by 3.8\% F1-score. 
`PN-GNN' denotes the GNN that employs the point cloud features as node features and utilizes the NCR. 
`PVN-GNN' means integrating the NCR with the `PV-GNN' model, which brings a performance gain of 0.6\%. 

\begin{table}[!t]
\centering 
\caption{Ablation study of different components used in our BCNet. `PF', `VF', and `NCR' denote point feature, voxel feature, and neighborhood consistency regularization, respectively. Note that all methods involve a voxel feature needed to train the UNet. \textit{p-value} is calculated between our model and other models under three different seeds.}
\begin{tabular}{@{}cccccccc@{}}
\toprule
\multirow{2}{*}{Methods} & \multirow{2}{*}{CNN} & \multicolumn{3}{c}{GNN}              & \multirow{2}{*}{F1-score}  & \multirow{2}{*}{p-value}\\ \cmidrule(lr){3-5}
                         &                      & PF         & VF         & NCR        &                         \\ \midrule
CNN                      & \checkmark           &            &            &            & $85.7_{\pm8.5}$        & \textless  0.001         \\
V-GNN                    &                      &            & \checkmark &            & $91.4_{\pm5.0}$        & 0.04        \\
P-GNN                    &                      & \checkmark &            &            & $88.2_{\pm5.7}$        & \textless  0.001        \\
VN-GNN                   &                      &            & \checkmark & \checkmark & $91.7_{\pm5.5}$        & 0.06 \\
PN-GNN                   &                      & \checkmark &            & \checkmark & $90.7_{\pm5.9}$        & 0.01 \\
PV-GNN                   &                      & \checkmark & \checkmark &            & $91.8_{\pm6.0}$        & 0.07 \\
PVN-GNN                  &                      & \checkmark & \checkmark & \checkmark & $92.4_{\pm6.5}$        & 0.16 \\
BCNet (GNN)              & \checkmark           & \checkmark & \checkmark & \checkmark & $91.2_{\pm6.3}$        & 0.03 \\
BCNet (CNN)              & \checkmark           & \checkmark & \checkmark & \checkmark & $93.7_{\pm5.5}$         & -\\ \bottomrule
\end{tabular}
\label{tab:ablacls}
\end{table}

`BCNet (GNN)' and `BCNet (CNN)' denote the results of the CNN (UNet) and GNN branches in our BCNet, respectively. Under the guidance of the GNN, the CNN branch in BCNet obtains 93.7\% F1-score and impressively outperforms the CNN baseline (85.7\%) by \textbf{8\% on F1-score}. The improvement well verifies the idea of structure-guided learning that training the GNN branch can boost the feature learning and classification result of the CNN branch. Of the reasons the performance of the GNN part degrades, one is that the GNN result needs to be recovered to the cube with an unlearnable region growth algorithm. The reason for PVN-GNN is better than BCNet (GNN) on F1-score is that PVN-GNN uses a pre-trained and fixed classification CNN network to extract static voxel features, which are used to initialize the nodes in PVN-GNN. In this situation, the voxel features are stable and discriminative for classification. In contrast, BCNet (GNN) uses voxel features extracted from a CNN which is updated in real-time, which may introduce unstable noises and negatively affect the initial training process of the GNN branch, leading to degraded performance. Note that `BCNet (GNN)' still achieves the state-of-the-art result among existing methods in Table~\ref{sota-cls}. Interestingly, during the inference, BCNet can run only the CNN branch without relying on the GNN branch and thus enjoys both the best performance 93.7\%, and the same efficiency as the CNN baseline.

\section{Discussion}

\subsection{Constrains and Limitations}
The one-stage framework utilizes UNet for direct multi-class segmentation from CT images and processes samples in approximately 3.5 seconds each. Despite its speed, the performance metrics show significant limitations, with an accuracy of only 36.1 (±15.7), precision of 34.3 (±12.2), and particularly low recall of 11.8 (±3.7), resulting in an F1-score of 17.0 (±4.9). The p-value \textless 0.001 highlights the considerable performance gap compared to the two-stage approach. 
Our two-stage framework, though more time-consuming, clearly outperforms the one-stage. In the training process, we start with UNet generating a binary mask in 3.5 seconds per sample, followed by 
construction of bronchial tree (1.5 seconds per sample) and CNN-GNN based classification (1.01 seconds per sample), totaling around 6 seconds per sample. This approach attains an accuracy of 93.6 (±5.6), precision of 93.4 (±6.1), and recall of 91.7 (±8.2), with an F1-score of 92.5 (±7.0). It is worth noting that during the inference stage, we only need 1 second to obtain the classification result from the CNN branch without prepossessing like building a bronchial tree.

\begin{table}[!h]
\centering
\caption{Comparison of One-Stage and Two-Stage Frameworks for CT Image Segmentation. The one-stage approach employs UNet for direct multi-class segmentation from CT images. The two-stage framework first utilizes UNet to generate a binary mask from the CT, followed by the application of BCNet for multi-class classification.}
\begin{tabular}{cccccc}
\toprule
Method &Accuracy  & Precision  & Recall  & F1-score  & p-value\\ \midrule
one-stage & 36.1$_{\pm15.7}$ & 34.3$_{\pm12.2}$ & 11.8$_{\pm3.7}$ & 17.0$_{\pm4.9}$ & \textless  0.001 \\
two-stage & 93.6$_{\pm5.6}$& 93.4$_{\pm6.1}$& 91.7$_{\pm8.2}$& 92.5$_{\pm7.0}$& - \\
\bottomrule
\end{tabular}
\label{tab:1v2}
\end{table}

The one-stage method's failure is largely due to its inability to separately optimize segmentation and classification, forcing UNet to manage both tasks simultaneously. This can lead to conflicts between segmentation accuracy and classification precision. In contrast, the two-stage method allows each network (UNet for segmentation and BCNet for classification) to specialize and optimize its respective task, resulting in significantly enhanced overall performance.
In summary, the two-stage framework, despite its longer processing time, offers a substantial increase in performance, making it preferable for applications where accuracy is critical. The one-stage method, although faster, suffers from significant performance deficits, rendering it less suitable for high-stakes clinical applications.

Another obstacle in our current work is the performance of the abnormal category predictions. Despite BCNet being the best-performing model with an accuracy of 61.2 and F1-score of 54.8, it, along with CNN and GIN, shows substantial room for improvement, particularly in recall rates. Future efforts will focus on refining these models through advanced neural architectures, increased data augmentation, sophisticated feature engineering, and hybrid model approaches. Moreover, implementing advanced training techniques such as transfer learning could also enhance our model's ability to accurately identify abnormal categories, addressing the critical need for reliable detection in medical imaging. These strategies are pivotal as we aim to close the performance gap between abnormal and normal category predictions.

\begin{table}[!h]
\centering
\caption{Analysis on the hyper-parameters $\alpha$ and $\beta$. Parameter $\alpha$ is used in the neighborhood consistency regularization (NCR) to balance the trade-off between cross-entropy loss and NCR, as shown in Eq.~\ref{eq:cls-NCR}. Parameter $\beta$ adjusts the trade-off between the cross-entropy loss on the CNN and GNN branches, as detailed in Eq.~\ref{eq:clsloss}.}
\begin{tabular}{@{}cccccc}
\toprule
$\alpha$& 0 & 0.5 & 1 & 1.5 & 2 \\
F1-score  & $92.8_{\pm6.1}$ & $93.1_{\pm5.8}$ & $93.7_{\pm5.5}$ & $93.3_{\pm4.9}$ & $93.0_{\pm6.5}$ \\
\midrule
$\beta$& - & 0.5 & 1 & 2 & -\\
F1-score  & - & $93.5_{\pm4.7}$ & $93.7_{\pm5.5}$ & $93.4_{\pm5.3}$ & -\\
\bottomrule
\end{tabular}
\label{tab:parameters}
\end{table}

\subsection{Sensitive and Qualitative Analysis}
The sensitivity analysis for our BCNet architecture, detailed in Table~\ref{tab:parameters}, evaluates the impact of hyper-parameters $\alpha$ and $\beta$. The parameter $\alpha$, which governs the balance between cross-entropy loss and neighborhood consistency regularization (NCR) as per Eq.~\ref{eq:cls-NCR}, shows that increasing $\alpha$ from 0 to 1 enhances the F1-score, reaching a peak at $\alpha=1$ ($93.7$). This indicates optimal NCR application, improving model generalization. Beyond this, further increases in $\alpha$ slightly degrade performance, suggesting over-regularization. 

Similarly, the parameter $\beta$, adjusting the cross-entropy loss balance between the CNN and GNN branches as outlined in Eq.~\ref{eq:clsloss}, reveals that both $\beta=0.5$ and $\beta=1$ optimize the model's performance with F1-scores of $93.5$ and $93.7$, respectively. A shift to $\beta=2$ results in a minor decrease in the F1-score to $93.4$, indicating less effective loss distribution.

\begin{table}[!h]
\centering
\caption{Analysis on the number of points $K$ in the graph construction process.}
\begin{tabular}{@{}cccccc@{}}
\toprule
$K$ & 6 & 8 & 10 & 12 & 14 \\ \midrule
F1-score & $93.3_{\pm5.8}$ & $93.4_{\pm5.9}$ & $93.7_{\pm5.5}$ & $93.2_{\pm5.2}$ & $93.5_{\pm6.1}$ \\ \bottomrule
\end{tabular}
\label{tab:k}
\end{table}
Table~\ref{tab:k} demonstrates the effect of varying $K$ on the F1 score. As $K$ increases from 6 to 10, the F1 score improves, reaching a peak of $93.7$ at $K=10$. However, further increases in $K$ result in marginal changes, suggesting an optimal $K$ range of 1 to 1.5 for achieving the highest F1 score. 

We also provide the qualitative comparison result for bronchus segmentation in Fig.~\ref{visualization}. As we can see, our method produces less misclassified bronchus (marked in red box) than other approaches. 

\section{Conclusion}
In this paper, we present the BCNet, a structure-guided learning framework for bronchus classification based on CT imaging data. 
With the simultaneous training of our proposed Point-Voxel GNN at the segment level, the BCNet learns to harvest structure-aware representations for the shared convolutional backbone and hence yields better predictions at the voxel-level branch. Besides, we exploit the structure prior with a novel neighborhood consistency regularization to boost the performance. 
We contribute the BronAtlas benchmark that contains 100 CT scans with voxel-wise masks and segment-level labels to facilitate future research. The experimental results on the BronAtlas benchmark show that our proposed model significantly exceeds the state-of-the-art methods.

\bibliographystyle{ieeetr}
\bibliography{ref}

\end{document}